\documentclass[10pt,twoside]{article}
\usepackage{fleqn,espcrc1}

\usepackage{graphicx}

\def\beq{\begin{equation}}
\def\eeq{\end{equation}}
\def\bea{\begin{eqnarray}}
\def\eea{\end{eqnarray}}
\def\bq{\begin{quote}}
\def\eq{\end{quote}}

\def\nnb{\nonumber}
\def\ga{\left(}
\def\dr{\right)}

\def\rar{\rightarrow}
\def\lrar{\Longrightarrow}

\def\nnb{\nonumber}
\def\la{\langle}
\def\ra{\rangle}
\def\nin{\noindent}
\def\ba{\begin{array}}
\def\ea{\end{array}}

\def\bl{\bullet}

\def\als{\alpha_s}

\def\g2{ \la\alpha_s G^2 \ra}
\def\g3{g^3f_{abc}\la G^aG^bG^c \ra}
\def\g4{\la\als^2G^4\ra}

\newcommand{\AmS}{{\protect\the\textfont2
  A\kern-.1667em\lower.5ex\hbox{M}\kern-.125emS}}

\title{On the Quark and Gluon Substructure of the $\sigma$ and other Scalar Mesons\thanks{Plenary
talk given at the Workshop on `` Possible existence of the Light $\sigma$-resonance and its
applications to hadron physics" (Yukawa Institute for Theoretical Physics, Kyoto, Japan, 11-14th June
2000) and Review talk given at the QCD 00 Euroconference (15th anniversary) Montpellier, France 6-13th July
2000.}}

\author{Stephan Narison\address{
Laboratoire de Physique Math\'ematique,
Universit\'e de Montpellier 2\\
Place Eug\`ene Bataillon,
34095 - Montpellier Cedex 05, France\\
E-mail:
narison@lpm.univ-montp2.fr}}%
        
\begin{document}

\maketitle

\begin{abstract}
Within the increasing experimental evidence of the existence of the $I=0$ scalar
low mass and wide $\sigma$ meson, we review the {\it first} analysis of its quark and gluon
substructure from QCD spectral sum rules and some low-energy theorems. The present
data favour equal components of $\bar uu+ \bar dd$ and of $gg$ in its wave function,
which make the wide $\sigma$ and the narrow $f_0(980)$ as $\eta'$-like particles.
A coherent picture of the other $I=0$ scalar mesons spectra within this mixing scheme is
shortly discussed. We also expect the $a_0(980)$ to be the lowest isovector $\bar ud$
state, and the $K^*_0(1430)$ its $\bar ds$ partner.
\end{abstract}

\section{INTRODUCTION}

\nin
Since the discovery of QCD, it has been emphasized  \cite{GM} that exotic mesons beyond the
standard octet, exist as a consequence of the non-perturbative aspects of
quantum chromodynamics (QCD). Among these states, the
$\eta'$ meson is a {\it peculiar} pseudoscalar state, as it should contain a large gluon
component (the 1st well-established gluonium!) in its wave function for explaining why its
mass is not degenerate with the one of the pion but only vanishes in the large $N_c$ limit
(solution of the so-called $U(1)_A$ problem)
\cite{U1}, though the
$\eta'$ likes to couple to
$\bar qq$ mesons in its decay (e.g. $\eta'\rar\eta\pi\pi$,...). This peculiar feature
has lead to certain confusion in the interpretation of its nature. Even, at present,
many physicists claim misleadingly that the $\eta'$ is a $\bar qq$ state, from the conclusion
based on the successful quark model prediction of its hadronic and $2\gamma$
couplings. This feature is not too surprising due to the manifestation of the
strong OZI-violation for low mass state in this channel. However, the $\eta'$ mass is lower than
the direct calculation of the pseudoscalar glueball mass from QCD spectral sum rules (QSSR)
and lattice QCD (see Table 1). This apparent discrepancy can be understood from the
 crucial r\^ole played by the $\eta'$ in the evaluation of the $U(1)_A$ topological
susceptiblity and of its slope
\cite{SHORE}. In the effective Lagrangian approach, it has been expected since the pioneer
work of Nambu-Lonasinio that an analogue of the pion of the non-linear
$\sigma$ model (NL$\sigma M$) exists in the linear $\sigma$  model (L$\sigma M$)
\cite{LSIGMA}, but this L$\sigma M$ has not attracted too much attention in the past because it
has been observed that the $SU(2)_L\times SU(2)_R$ chiral symmetry of QCD is non-linearly
realized (this has made the success of the chiral perturbation theory (CHpT) approach
\cite{NLSIGMA}). It also occurs that the Lagrangian of the $L\sigma M$ is not unique such
that no definite predictions from QCD first principles can be done. In addition, the inclusion of
the resonances into the effective Lagrangian is not completly settled \cite{RAFEL}.\\
However, this analogy between the pion and the $\sigma$ meson may not be very appropriate
as they are particles associated to symmetries of a different nature ($SU(2)_L\times
SU(2)_R$ for the pion and
$U(1)_V$ for the $\sigma$). Instead, a comparison of the
$\sigma$ with the
$\eta'$ meson looks more valuable within a chiral $U(3)_A\times U(3)_V$ Lagrangian
\cite{U1,U11}, as both particles are associated respectively to the U(1) axial and vector
symmetries, where in terms of the quark and gluon fields, the corresponding QCD currents
are respectively, the
$U(1)_A$ anomaly (divergence of the singlet axial current):
\bea
\partial_\mu A^\mu(x)&=&\ga \frac{\alpha_s}{8\pi}\dr
\mbox{tr}~G_{\alpha\beta}
\tilde{G}^{\alpha\beta}+\sum_{u,d,s}m_q\bar
q(i\gamma_5)q~,~~~~~~
\eea
and the dilaton current:
\bea
\theta_{\mu}^{\mu}&=& \frac{1}{4}\beta(\alpha_s)
G_{\alpha\beta}G^{\alpha\beta}+\ga 1+\gamma_m(\alpha_s)\dr\sum_{u,d,s}m_q\bar qq~,
\eea
which is the
trace of the energy-momentum tensor $\theta_{\mu\nu}$.
The sum over colour is understood; $q$,~ $G_{\mu\nu}$ and $\tilde{G}_{\mu\nu}$ are respectively the
quark, the gluon field strength and its dual; $m_q$ is the light quark
 running mass;
$\beta$ and $\gamma_m$ are respectively the $\beta$-function and quark mass anomalous
dimension. 
\\ In this
talk, we shall  discuss the $\sigma$ meson from this point of view of the analogy with the
$\eta'$-meson by using QCD spectral sum rules (QSSR) 
\`a la SVZ \cite{SVZ} (for a review, see e.g.: \cite{SNB}) and some low-energy theorems based on
Ward identities. The discussions are based on the works in \cite{VEN,BRAMON2,SNG}.
\footnote{Discussions of the $\bar qq$ and/or gluonium nature of scalar states can
also be found in e.g.\cite{NSVZ,CHAN,MINK,ELLIS,MONT,GAST,BUGG}.}.
\section{AN OUTLINE OF QCD SPECTRAL SUM RULES (QSSR)}
\subsection{Description of the method}
Since its discovery in 79, QSSR has proved to be a
powerful method for understanding the hadronic properties in terms of the
fundamental QCD parameters such as the QCD coupling $\alpha_s$, the (running)
quark masses and the quark and/or gluon QCD vacuum condensates.
The description of the method has been often discussed in the literature,
where a pedagogical introduction can be, for instance, found in the book \cite{SNB}. In
practice (like also the lattice), one starts the analysis from the two-point correlator:
\beq
\psi_H(q^2) \equiv i \int d^4x ~e^{iqx} \
\la 0\vert {\cal T}
J_H(x)
\ga J_H(0)\dr ^\dagger \vert 0 \ra ~,
\eeq
built from the hadronic local currents $J_H(x)$, which select some specific quantum numbers.
However, unlike the lattice which evaluates the correlator in the Minkowski space-time,
one exploits, in the sum rule approaches, the analyticity property of the
correlator which obeys the well-known K\"allen--Lehmann dispersion relation:
\beq
\psi_H (q^2) = 
\int_{0}^{\infty} \frac{dt}{t-q^2-i\epsilon}
~\frac{1}{\pi}~\mbox{Im}  \psi_H(t) ~ + ...,
\eeq
where ... represent subtraction points, which are
polynomials in the $q^2$-variable. In this way, the $sum~rule$
expresses in a clear way the {\it duality} between the integral involving the 
spectral function Im$ \psi_H(t)$ (which can be measured experimentally), 
and the full correlator $\psi_H(q^2)$. The latter 
can be calculated directly in the
QCD Euclidean space-time using perturbation theory (provided  that
$-q^2+m^2$ ($m$ being the quark mass) is much greater than $\Lambda^2$), and the Wilson
expansion in terms of the increasing dimensions of the quark and/or gluon condensates which
 simulate the non-perturbative effects of QCD. 

\subsection{Beyond the usual SVZ expansion}

Using the Operator Product Expansion (OPE) \cite{SVZ}, the two-point
correlator reads:
\beq
\psi_H(q^2)
\simeq \sum_{D=0,2,4,...}\frac{1}{\ga -q^2 \dr^{D/2}} 
\sum_{dim O=D} C(q^2,\nu)\la {\cal O}(\nu)\ra~,
\eeq
where $\nu$ is an arbitrary scale that separates the long- and
short-distance dynamics; $C$ are the Wilson coefficients calculable
in perturbative QCD by means of Feynman diagrams techniques; $\la {\cal O}(\nu)\ra$
are the quark and/or gluon condensates of dimension $D$.
In this paper, we work in the massless quark limit. Then, one may expect
the absence of the terms of dimension 2 due to gauge invariance. However, it has been
emphasized recently \cite{ZAK} that  the resummation of the large order terms of the
perturbative series, and the effects of the higher dimension condensates due e.g. to instantons, can
be mimiced by the effect of a tachyonic gluon mass, which might be understood
from the short distance linear part of the QCD potential. The strength of
this short distance mass has been estimated from the $e^+e^-$ data to be
\cite{SNI,CNZ}:
$
\frac{\alpha_s}{\pi}\lambda^2\simeq -(0.06\sim 0.07) ~\rm{ GeV}^2,
$
which leads to the value of the square of the (short distance) string tension:
$
\sigma \simeq -\frac{2}{3}{\alpha_s}\lambda^2\simeq [(400\pm 20)~\rm{ MeV}]^2
$
in an (unexpected) good agreement with the lattice result \cite{TEPER} of about
$[(440\pm 38)~\rm{ MeV}]^2$.
The strengths of the vacuum condensates having dimensions $D\leq 6$ are also under
good control: namely $2m\la\bar qq\ra =-m^2_\pi f^2_\pi$ from pion PCAC,
$\la\alpha_s G^2\ra =(0.07\pm 0.01)$ GeV$^2$ from $e^+e^-\rar I=1$ data \cite{SNI} and from 
the heavy quark mass-splittings \cite{SNH}, $\alpha_s  \la\bar qq\ra^2\simeq
5.8 \times 10^{-4}$ GeV$^6$ \cite{SNI}, and $g^3\la G^3\ra\approx$ 1.2 GeV$^2\la\alpha_s G^2\ra$ from
dilute gaz instantons \cite{NSVZ}.
\subsection{Spectral function}
The spectral function is often parametrized
using the ``na\"{\i}ve" duality ansatz:
\beq
\frac{1}{\pi}~\mbox{Im}  \psi_H(t)\simeq 2M_H^{2n}f_H^2 \delta (t-M_H^2)+ \rm{``QCD
~continuum"}
\times \theta(t-t_c)~, 
\eeq
which has been tested \cite{SNB} using $e^+e^-,~\tau$-decay data, to give a good description of the
spectral integral in the sum rule analysis; $f_H$ (analogue to $f_\pi$) is the
the hadron's coupling to the current ; $2n$ is the dimension of the
correlator, while $t_c$ is the QCD continuum's threshold. 

\subsection{Form of the sum rules and optimization procedure}
Among the different sum rules discussed in the literature within QCD \cite{SNB} (Finite
Energy Sum rule (FESR) \cite{RAFAEL}, $\tau$-like sum rules \cite{BNP},...), we shall mainly be
concerned here with:\\
$\bullet$ The exponential Laplace unsubtracted sum rule (USR)
and its ratio:
\beq\label{usr}
{\cal L}_n(\tau)
= \int_{0}^{\infty} {dt}~t^n~\mbox{exp}(-t\tau)
~\frac{1}{\pi}~\mbox{Im} \psi_H(t)~,~~~~~~~~~~{\cal R}_{n} \equiv -\frac{d}{d\tau} \log {{\cal
L}_n}~,~~~~~~~(n\geq 0)~;
\eeq
$\bullet$ The subtracted sum rule (SSR):
\beq\label{ssr}
{\cal L}_{-1}(\tau)
= \int_{0}^{\infty} \frac{dt}{t}~\mbox{exp}(-t\tau)
~\frac{1}{\pi}~\mbox{Im} \psi_H(t) +\psi_H(0)~.
\eeq
The advantage of the Laplace sum 
rules with respect to the previous dispersion relation is the
presence of the exponential weight factor, which enhances the 
contribution of the lowest resonance and low-energy region
accessible experimentally. For the QCD side, this procedure has
eliminated the ambiguity carried by subtraction constants,
arbitrary polynomial
in $q^2$, and has improved the convergence of
the OPE by the presence of the factorial dumping factor for each
condensates of given dimensions. 
The ratio of the sum rules is a useful quantity to work with,
 in the determination of the resonance mass, as it is equal to the 
meson mass squared, in the usual duality ansatz parametrization.
As one can notice, there are ``a priori" two free external parameters $(\tau,
t_c)$ in the analysis. The optimized result will be (in principle) insensitive
to their variations. In some cases, the $t_c$-stability is not reached due to the
too na\"{\i}ve parametrization of the spectral function. One can either fixed the 
$t_c$-values by the help of FESR (local duality) or improve the
parametrization of the spectral function by introducing threshold effects fixed by
chiral perturbation theory, ..., in order to restore the $t_c$-stability of the
results. The results discussed below satisfy these stability criteria.
\section{UNMIXED GLUONIA MASSES AND DECAY CONSTANTS}
Before discussing the specific scalar channel, let's present the situation of gluonia/glueball
mass calculations as a guide for the forthcoming discussions.
\begin{table}[hbt]
\caption{ Unmixed gluonia masses and decay constants from QSSR \cite{SNG} compared with the
lattice.  } 
\begin{center}
\begin{tabular}[h]{ccccccc}
\hline 
$J^{PC}$& Name&\multicolumn{3}{l} {Mass [GeV]
}& $\sqrt{t_c}$ [GeV]&$f_H$ [MeV]\\
\cline{3-5}
&&Estimate& Upper Bound&Lattice \cite{TEPER,LATT} &&  \\
\hline 
 $0^{++}$&$G$&$ 1.5\pm 0.2$& $2.16\pm 0.22$&$1.60\pm
0.16$&$2.1$&$390\pm 145$\\
&$3G$& 3.1&3.7&&3.4&62\\
$2^{++}$&$T$&$2.0\pm 0.1$&$2.7\pm 0.4$&$2.26\pm
0.22$&$2.2$&$80\pm 14$ \\ 
$0^{-+}$&$P$&$2.05\pm 0.19$&$2.34\pm 0.42$&$2.19\pm
0.32$&$2.2$&$8
\sim17$ \\
\hline 
\end{tabular}
\end{center}
\end{table}
\subsection{The currents}
In addition to the pseudoscalar and scalar currents introduced previously, we shall deal with
the tensor and 3-gluon currents (standard notations):\bea
\theta_{\mu\nu}=-G_{\mu}^{\alpha}G_{\nu\alpha}+\frac{1}{4}g_{\mu\nu}~,~~~~~~~~
J_3=g f_{abc} G^a_{\alpha\beta}G^b_{\beta,\gamma}G^c_{\gamma\alpha}~,
\eea
\subsection{Masses and decay constants}
The unmixed gluonia masses from the unsubtracted QCD Spectral Sum Rules
(USR) \cite{SNG} are compared in Table 1 with the ones from the lattice
\cite{TEPER,LATT} in the quenched approximation, where we use the
conservative guessed estimate of about
15\% for the different lattice systematic errors  (separation  of the lowest ground
states from the radial excitations, which are expected to be nearby as
indicated by the sum rule analysis; discretisation; quenched
approximation,...). 
One can notice an excellent agreement between the USR 
 and the lattice results, with the mass hierarchy:
$M_{0^{++}}\leq M_{0^{-+}}\approx M_{2^{++}}$,  expected from some QCD inequalities \cite{WEST}.
However, this is not the whole story ! Indeed, one can notice that in the pseudoscalar channel, the
predicted value of the mass of the $0^{-+}$ is too high compared with the mass of the $\eta'$, which is
not surprising for the reasons explained  in the introduction. We shall see in the next section that
the same phenomena occur for the scalar channel.
\section{UNMIXED SCALAR GLUONIA}
\subsection{The need for a low mass $\sigma_B$ from the sum rules}
Using the mass and decay constant of the scalar gluonium $G$ in Table 1 from
the USR (Eq.\ref{usr}), into the SSR (Eq.\ref{ssr}) sum rules, where \cite{NSVZ} $\psi_s(0)\simeq
-16(\beta_1/\pi)\la
\alpha_s G^2\ra$, one can notice \cite{VEN,SNG} that one needs a low mass resonance $\sigma_B$ 
for a consistency of
the two sum rules. Using, e.g., $M_{\sigma_B}\simeq 1$ GeV \footnote{We cannot fix both the mass and
decay constant of the $\sigma$ from the two sum rules. We give in the original papers a more complete
analysis for different values of the $\sigma$ mass ranging from 500 to 1 GeV. }, one gets
\cite{VEN,SNG}:
$
f_{\sigma_B}\approx 1 ~\rm{GeV}
$, which is larger than $f_G\simeq$ .4 GeV.
\subsection{Low-energy theorems (LET) for the couplings to meson pairs} 
In order to estimate the couplings of the gluonium to meson pairs, we use some sets of low-energy theorems (LET)
based on Ward identities for the vertex:
\beq
V(q^2\equiv (p-p')^2=0)\equiv \la H(p)|\theta_{\mu}^{\mu}|H(p')\ra\simeq 2m^2_H~,~~ ~~~~\rm{and}~~~~~~ V'(0)=1~,
\eeq
and write the vertex in a dispersive form. $H$ can be a Goldstone boson ($\pi,K,\eta_8$), a $\eta_1$-
$U(1)_A$-singlet~, or a $\sigma_B$. Then, one obtains the sum rules for the hadronic couplings:
\beq
\frac{1}{4}\sum_{\sigma_B, \sigma'_B, G}g_{SHH}\sqrt{2}f_S\approx 2M^2_H~,~~~~~~~~~~~~~~~~~~~~~~~
\frac{1}{4}\sum_{\sigma_B, \sigma'_B, G}g_{SHH}\sqrt{2}f_S/M^2_S\simeq 1~.
\eeq
$\bl$ {\bf Decays into $\pi\pi$}: Neglecting, to a first approximation the $G$-contribution, the $\sigma_B$
and
$\sigma'_B$ widths to
$\pi\pi,~KK,...$
 (we take $M_{\sigma'}\approx
1.37$ GeV as an illustration) are \cite{SNG} \footnote{However, one can notice
that the $\sigma$ coupling to pion pairs decreases like $1/f_\sigma$, i.e., a too low value of the mass
say less than 500 MeV, leads to a width less than 100 MeV, not favoured by the present data.}:
\beq
\Gamma(\sigma_B\rar\pi\pi)\approx 0.8~{\rm GeV}~,~~~~~~~~~~~~~~~~~~~~~~
\Gamma(\sigma'_B\rar\pi\pi)\approx 2~{\rm GeV}~,
\eeq
which suggests a huge OZI violation and seriously questions the validity of the
lattice results in the quenched approximation. Similar conclusions have been reached in
\cite{MINK,ELLIS,MONT,GAST}. For testing the above
result, one should evaluate on the lattice, the decay mixing 3-point function $V(0)$ responsible for such decays
using dynamical fermions. \\
$\bl$ {\bf Decays into $\eta'\eta$ and $\eta\eta$}: Using $\eta'\approx cos\theta_P\eta_1$ and
$\eta\approx sin\theta_P\eta_1$, where $\theta_P$ is the pseudoscalar mixing angle, the previous
LET implies the {\it characteristic gluonium decay} (we use $M_G\approx 1.5$ GeV and assume a
G-dominance in the sum rule) \cite{VEN,SNG}:
\beq
\Gamma(G\rar\eta\eta')\approx (5-10)~{\rm MeV}~,~~~~~~
\frac{\Gamma(G\rar\eta\eta)}{\Gamma(G\rar\eta\eta')}\approx 0.22 ~~~:
~~~g_{G\eta\eta}\simeq \sin\theta_P~g_{G\eta\eta'}~.
\eeq
$\bl$ {\bf Decay into $4\pi^0$}: Assuming that the $G$ decay into $4\pi^0$ occurs through
$\sigma_B\sigma_B$, and using the data for $f_0(1.37) \rar (4\pi^0)_S$, one obtains from the
previous sum rule \cite{VEN,SNG}:
\beq\Gamma(G\rar\sigma_B\sigma_B\rar 4\pi)\approx (60-140)~{\rm MeV}~.
\eeq
\subsection{$\gamma\gamma$ widths and $J/\psi\rar\gamma S$  radiative decays}
These widths can be estimated from the quark box or anomaly diagrams \cite{VEN,SNG}. 
The $ \gamma\gamma$ widths
of the $\sigma, \sigma'$ and $G$ are much smaller (factor 2 to 5) than
$\Gamma(\eta'\rar\gamma\gamma) \simeq$ 4 keV, while
$B (J/\psi\rar \gamma$ $\sigma, \sigma'$ and $G$) is about 
10 times smaller than 
$B(J/\psi\rar \gamma \eta')\approx 4~
10^{-3}$. These are typical values of gluonia widths and production rates \cite{CHAN}.
The absence of the $\sigma$ in $\gamma\gamma$ scattering and its presence
in $J/\Psi$ radiative decays \cite{GAST} are a
strong indication of its large gluon component.
\section{UNMIXED SCALAR QUARKONIA}
\subsection{The $a_0(980)$}
The  $a_0(980)$ is the most natural meson candidate associated to the divergence of the vector
current:
$
\partial_\mu V^\mu (x)\equiv (m_u-m_d) \bar u(i\gamma_5) d.
$
Previous different sum rule analysis of the associated two-point
correlator gives \cite{SNB}:
$
M_{a_0}\simeq 1~\rm{ GeV}$ and the conservative range $ f_{a_0}\simeq (0.5-1.6) ~{\mbox
MeV}~~(f_\pi=93~{\mbox MeV}), $ 
in agreement with the value 1.8 MeV from a hadronic kaon tadpole mass difference approach plus
a $a_0$ dominance of the $K\bar K$ form factor. A
3-point function sum rule analysis gives the widths \cite{BRAMON2,SN4,SNG}:
\beq
\Gamma (a_0\rar\eta\pi) \simeq 37~ {\rm MeV}~,~~~~~~~~~~~~~~~~~~~~~~~~\Gamma (a_0\rar\gamma\gamma)
\simeq (0.3-1.5)~{\rm keV}~,
\eeq
while from $SU(3)$ symmetry, we expect to have:
$
g_{a_0 K^+K^0}\simeq \sqrt{
\frac{3}{2}} g_{a_0\eta\pi}.
$
Analogous sum rule  analysis in the four-quark scheme \cite{SN4,SNB} gives similar values of the masses
and hadronic couplings but implies a too small value of the $\gamma\gamma$ width
due to the standard QCD loop-diagram factor suppressions.
The $(\bar uu-\bar dd)$ quark assignement for the $a_0(980)$ is supported by present data
and alternative approaches \cite{MONT,GAST,BUGG}. 
\subsection{The isoscalar partner $S_2\equiv \bar uu+\bar dd$ of the $a_0(980)$}
Analogous analysis of the corresponding 2-point correlator gives $M_{S_2}\approx M_{a_0}$
as expected from a good $SU(2)$ symmetry, while using 3-point function and $SU(3)$ relation, its widths are
estimated to be
\cite{SNG,BRAMON2}:
\beq
\Gamma (S_2\rar\pi^+\pi^-) \simeq 120~ {\rm MeV}~,~~~~~~~~~~~~~~~
\Gamma (S_2\rar\gamma\gamma)\simeq \frac{25}{9}\Gamma (a_0\rar\gamma\gamma) \simeq
0.7~{\rm keV}~.
\eeq
\subsection{The $K^*_0(1430)\equiv \bar ds$ and $S_3\equiv \bar ss$ states}
An analysis of the $K^*_0$-$a_0$ mass shift due to $SU(3)$
breakings (strange quark mass and condensate) \cite{SNB,SNG} fixes the mass of the $K^*_0$ to be
around 1430 MeV. If a candidate around 900 MeV is confirmed, it will then be hard to reconcile with
a $\bar qq$ structure. An
analysis of the
$S_3$ over the
$K^*_0$ 2-point functions gives
\cite{SNG}:
\beq
M_{S_3}/M_{K^*_0}\simeq 1.03\pm 0.02 ~~\lrar~~~M_{S_3}\simeq 1474~{\rm MeV}~,~~~~~~~~f_{S_3}\simeq (43\pm
19)~\rm{MeV}~,\eeq in agreement with the lattice result \cite{LEE}, while the 3-point function leads to
\cite{SNG}:
\beq
\Gamma (S_3\rar K^+K^-)\simeq (73\pm 27)~{\rm MeV}~,~~~~~~~~~~~~
 \Gamma (S_3\rar\gamma\gamma) 
\simeq 0.4~{\rm keV}~.
\eeq
In the usual sum rule approach (absence of large violations of the OPE at the sum rule stability
points), one expects a small mixing between the $S_2$ and $S_3$ mesons before the mixing with the
gluonium $\sigma_B$. 
\subsection{Radial excitations}
The propreties of the radial excitations cannot be obtained accurately from the sum rule
approach, as they are part of the QCD continuum which effects are minimized in the analysis.
However, as a crude approximation and using the sum rule results from the well-known channels ($\rho$,...),
one may expect that the value of $\sqrt{t_c}$ can localize approximately the position of the first radial
excitations. Using this result and some standard phenomenological arguments on the estimate of the
couplings, one may expect \cite{SNG}:
\bea
M_{S'_2}\approx 1.3~\rm{GeV},~~\Gamma (S'_2\rar \pi^+\pi^-)\approx (300\pm
150)~{\rm MeV},~~
 \Gamma (S'_2\rar\gamma\gamma)\approx (4\pm 2)~{\rm keV}~,\nnb\\
M_{S'_3}\approx  1.7~{\rm GeV},~~
\Gamma (S'_3\rar K^+K^-)\approx (112\pm 50)~{\rm
MeV},~~
 \Gamma (S'_3\rar\gamma\gamma)\approx (1\pm .5)~{\rm keV}.
\eea
\subsection{We conclude that:}
\nin
$\bullet$ Unmixed scalar quarkonia ground states are not wide, which excludes the interpretation
of the low mass broad $\sigma$ for being an ordinary $\bar qq$ state.\\
$\bullet$ There can be many states in the region  around 1.3 GeV ($\sigma', ~S_3, ~S'_2$),which 
should mix non-trivially in order to give the observed $f_0(1.37)$ and $f_0(1.5)$
states (see next sections).\\
$\bullet$ The $f_J(1.7)$ seen to decay mainly into $\bar KK$ \cite{SING}, if it is
confirmed to be a $0^{++}$ state, can be {\it the first radial excitation of the $S_3\equiv\bar ss$ state}, but
{\it definitely not the pure gluonium} advocated in \cite{WEIN2}.
\section{SCALAR MIXING-OLOGY}
\subsection{Mixing below 1 GeV and the nature of the $\sigma$ and $f_0(980)$}
In so doing, we consider the two-component mixing scheme of the bare states
$(\sigma_B,~S_2)$:
\beq
|f_0>\equiv -\sin\theta_s|\sigma_B>+\cos\theta_s|S_2>~,~~~~~~~~~~~~~
|\sigma>\equiv ~~~\cos\theta_s|\sigma_B>+\sin\theta_s|S_2>
\eeq
A sum rule analysis of the off-diagonal 2-point correlator \cite{MENES,SNB,SNG}:
\beq\label{offs}
\psi_{qg}^S(q^2) \equiv i \int d^4x ~e^{iqx}
\la 0\vert {\cal T}
\beta(\alpha_s)
G^2(x)\sum_{u,d,s}m_q\bar
qq~(0) \vert 0 \ra ,
\eeq
responsible for the mass-shift of the mixed
states gives a small {\it mass mixing angle} of about $15^0$, which has been confirmed by lattice calculations
using different input for the masses \cite{WEIN2} and from the low-energy theorems based on Ward
identities of broken scale invariance \cite{ELLIS}, if one uses there the new input values
\cite{SNB,SNI} of the quark and gluon condensates. In order to have more complete discussions on the
gluon content of the different states, one should also determine the {\it decay mixing angle}.  In so
doing, we use the predictions for
$\sigma_B, ~S_2\rar
\gamma\gamma$ obtained in the previous sections and the data
$\Gamma(f_0\rar\gamma\gamma)\approx 0.3$ keV. Then, we deduce  a {\it maximal decay mixing angle}
and the widths \cite{BRAMON2,SNG}:
\bea
|\theta_s| &\simeq& (40-45)^0~,\nnb\\
\Gamma (f_0\rar\pi^+\pi^-)&\leq& 134~ {\rm
MeV}~,~~~~~~~~~~~~~~~~~~~ g_{f_0K^+K^-}/g_{f_0\pi^+\pi^-}\approx 2~,\nnb\\
\Gamma (\sigma\rar\pi^+\pi^-)&\approx& (300-700)~ {\rm MeV}~,~~~~~~~~
\Gamma (\sigma\rar\gamma\gamma)\approx ~(0.2-0.5) ~{\rm keV}~.
\eea
The huge coupling of the $f_0$ to $\bar KK$ comes from the large mixing with the 
$\sigma$. For this reason, the $f_0$ can have a large singlet component, as also suggested from
independent analysis \cite{MONT,MINK}. Extending the previous $J/\psi \rar \gamma +X$ analysis into the
case of the $\phi$, one obtains the {\it new result} within this scheme \cite{SNU}:
\beq
Br [\phi\rar \gamma
+f_0(980)]\approx 1.3\times 10^{-4}~,
\eeq
in good agreement with the Novosibirsk data of $(1.93\pm 0.46\pm 0.5)\times 10^{-4}$.
\subsection{Mixing above 1 GeV  and nature of the $f_0(1.37)$,
$f_0(1.5)$, $f_0(1.6)$  and $f_0(1.7)$}
As already mentioned previously, this region is quite complicated due to the
proliferation of states.
Many scenarios have been proposed in the literature for trying to interpret this region
\cite{MONT,MINK,AMSLER,WEIN2}. However, one needs to clarify and to confirm the data \cite{GAST} for
selecting these different interpretations. We shall give below {\it some selection rules} which
can already eliminate some of these different schemes:\\
$\bl$ {\bf The $f_0(1.37)$}: If its decay into $\sigma\sigma\rar (4\pi^0)_S$, it signals mixings
with the
$\sigma$, $\sigma'$ and $G$. \\
$\bl$ 
{\bf The $f_0(1.5)$}:\\ 
$-$ If its decays into $\sigma\sigma$ and $\eta'\eta$, this signals a 
gluon component.\\
$-$ If its also couples to $\pi\pi$ and $\bar KK$, this signals a $\bar qq$ component which
may come from the $S'_2,~S_3$. 
Then, it can result from the $\bar qq$ mixings with the $\sigma,~\sigma'$ and
$G$,  like the
$f_0(1.37)$.\\  
$-$ If it couples weakly to $\pi\pi$ and $\bar KK$, while the ratio of its
$\eta\eta$ and $\eta'\eta$ is proportionnal to $1/\sin^2 \theta_P$, then it can be
an almost pure gluonium state, which can be identified with the $G$ in Table 1 obtained in the quenched
approximation (this approximation is expected to better at higher energies using $1/N_c$ arguments 
\cite{VEN,GAB}). \\
$\bl$ {\bf The $f_0(1.7)$}: If its decays mainly into $\bar KK$ \cite{GAST}, it is likely the radial
excitation
$S'_3$ of the
$S_3(\bar ss)$ state.
\section{CONCLUSIONS} 
We have seen that there is an increasing experimental evidence for the existence of the
$\sigma$ and other scalar mesons. QCD spectral sum rule (QSSR) and low-energy theorem (LET) provide a
first analysis of their gluon and quark substructure beyond the usual effective Lagrangian approach.
Present data favour a maximal quarkonium-gluonium mixing scheme for the $\sigma$-meson and for higher
mass scalar states. Within the scheme, the wide $\sigma$ and the narrow $f_0(980)$ appear to be
$\eta'$-like particles. The $a_0(980)$ is a $\bar uq$ state, while its strange partner $\bar ds$ is
expected to be the $K^*_0(1430)$ of PDG.
\section{ACKNOWLEDGEMENTS} It is a pleasure to thank the organizers of this
conference for their invitation to present this talk and the KEK theory group for a financial support.

\end{document}